%% file: 00-main.tex
\documentclass[sigconf]{acmart}
\usepackage[utf8]{inputenc}
\usepackage[T1]{fontenc}
\usepackage[inline]{enumitem}
\usepackage{booktabs}
\usepackage{amsmath}
\usepackage{amsfonts}
\usepackage{amssymb}
\usepackage{enumitem}
\usepackage{mathbbol}
\usepackage{mathtools}
\usepackage{hyperref}
\usepackage{url}
\usepackage{nicefrac}
\usepackage{microtype}
\usepackage{natbib}
\usepackage{multirow}
\usepackage{graphicx}
\usepackage{subfig}
\usepackage{etaremune}

\newcommand{\eg}{\emph{e.g.}}

\AtBeginDocument{%
  \providecommand\BibTeX{{%
    \normalfont B\kern-0.5em{\scshape i\kern-0.25em b}\kern-0.8em\TeX}}}

\setcopyright{acmcopyright}
\copyrightyear{2020}
\acmYear{2020}
\acmDOI{10.1145/1122445.1122456}
\acmConference[SIGIR '20]{SIGIR '20: ACM SIGIR Conference on Research and Development in Information Retrieval}{July 25--30, 2020}{Xi'an, China}
\acmBooktitle{SIGIR '20: ACM SIGIR Conference on Research and Development in Information Retrieval,
  July 25--30, 2020, Xi'an, China}
\acmPrice{15.00}
\acmISBN{978-1-4503-XXXX-X/18/06}

\newif\ifanon
\anonfalse

\begin{document}

\title{On the Reliability of Test Collections for Evaluating Systems of Different Types}

\ifanon
\author{Anonymous}
\else
\author{Emine Yilmaz}
\email{emine.yilmaz@ucl.ac.uk}
\affiliation{%
  \institution{University College London}
  \city{London}
  \country{UK}
}
\author{Nick Craswell}
\email{nickcr@microsoft.com}
\affiliation{%
  \institution{Microsoft}
  \city{Redmond}
  \country{USA}
}
\author{Bhaskar Mitra}
\email{bmitra@microsoft.com}
\affiliation{%
  \institution{Microsoft, University College London}
  \city{Montr\'eal}
  \country{Canada}
}
\author{Daniel Campos}
\email{dacamp@microsoft.com}
\affiliation{%
 \institution{Microsoft, University of Washington}
  \city{Redmond}
  \country{USA}
}
\fi

\input{01-abstract.tex}



\maketitle

\input{02-intro.tex}
\input{03-related.tex}
\input{05-experiment.tex}

\input{06-data.tex}

\input{07-result.tex}

\input{09-conclusion.tex}

\balance
\bibliographystyle{ACM-Reference-Format}
\begin{scriptsize}
\bibliography{bibtex}
\end{scriptsize}

\end{document}

%% file: 01-abstract.tex
\begin{abstract}
As deep learning based models are increasingly being used for information retrieval (IR), a major challenge is to ensure the availability of test collections for measuring their quality.
Test collections are generated based on pooling results of various retrieval systems, but until recently this did not include deep learning systems.
This raises a major challenge for reusable evaluation: Since deep learning based models use external resources (e.g. word embeddings) and advanced representations as opposed to traditional methods that are mainly based on lexical similarity, they may return different types of relevant document that were not identified in the original pooling.
If so, test collections constructed using traditional methods are likely to lead to biased and unfair evaluation results for deep learning (neural) systems.
This paper uses simulated pooling to test the fairness and reusability of test collections, showing that pooling based on traditional systems only can lead to biased evaluation of deep learning systems.
\end{abstract}

%% file: 02-intro.tex
\section{Introduction}
\label{sec:intro}

In recent years, deep neural models achieved state-of-the-art performance on a variety of tasks, and this has happened in a variety of fields ranging from computer vision to information retrieval (IR). It took relatively longer to observe such advances in core IR problems such as ranking~\citep{Dehghani17}, especially if we exclude results based on proprietary data---e.g., \citep{mitra2017learning}. Two possible explanations for this delay are related to training data and test data: 1) The lack of large-scale training datasets with tens or hundreds of thousands of queries, since large data would seem to be a requirement based on the experience in other fields, and 2) The lack of test collections to evaluate the quality of neural models in a fair and reliable manner. 

The TREC 2019 Deep Learning Track~\citep{craswell2020overview} addressed these problems by releasing large-scale training data, as well as by developing reliable and reusable test collections for evaluating the quality of various algorithms (ranging from traditional retrieval models such as BM25 to various neural models). The results of the track showed that when sufficient training data is available, most neural models tend to outperform the traditional retrieval models. 

Findings of the track were based on test collections that were created using depth-10 pools of both neural and traditional models. Inclusion of neural runs in pooling is highly unusual, since most test collections were created in the years before neural models had been developed, and even those developed more recently (e.g., \citep{allan2017trec}) did not have neural models trained on the large labeled datasets that were introduced in TREC 2019. 

Even though reusability of test collections for evaluating the quality of unseen systems has been widely studied in literature~\citep{Voorhees18, cormack1998efficient, Voorhees18, carterette2006minimal, sanderson2004forming, Sanderson05}, no previous work has analysed the reusability of test collections when they are created solely using systems of a particular type (e.g. traditional systems based on BM25, language modelling, etc.) towards evaluating the quality of systems that are of different type (e.g. neural systems based on deep learning models), which is the main question we aim to answer in this paper. Our approach is to simulate the earlier test collections, where pooling was with one type of model, to see whether this creates a bias against the other type of model, both when comparing within type and across types. 


Our results demonstrate that evaluation results obtained using test collections that are created solely using pools of traditional systems are less reliable in terms of evaluating the quality of neural systems.  Our findings suggest that such test collections should be used wit caution when evaluating the quality of neural systems as they may lead to incorrect conclusions regarding how the quality of a neural model compares with a traditional model, as well as how the neural model compares with another baseline neural model.


%% file: 03-related.tex
\section{Related work}
\label{sec:related}
A significant amount of research has been devoted to analysing the fairness and reusability of test collections for retrieval evaluation, where fairness refers to collection being unbiased in its evaluation to different runs that contributed to the construction of the pool and reusability refers to the fairness of the test collection towards evaluating the quality of the runs that did not contribute to the construction of the test collection~\citep{Voorhees18}. 

Zobel et al.~\citep{zobel1998reliable} argued that test collections constructed using depth-k pooling~\citep{sparck1975report} tend to be reasonably reusable and tend to be fair towards evaluating the quality of new systems. Various methods have been proposed in order to generate fair and reusable test collections with limited relevance labels~\citep{cormack1998efficient, Voorhees18, carterette2006minimal, sanderson2004forming, Sanderson05}.
Previous work has shown that when test collections are constructed using pools that are too small compared to the document collection size, the resulting pools could exhibit some bias (in particular, bias towards systems that retrieve documents that contain topic title words)~\citep{Buckley07}. 

While most of this previous work analysed the reusability of test collections in terms of their fairness towards evaluating new systems that did not contribute to the pool, none of the previous work analysed the reusability of such collections when they are constructed solely using systems that are of particular type (e.g. traditional systems) but are used to evaluate the quality of systems that are of a different type (e.g. neural systems). 

The TREC 2017 Common Core Track showed some evidence of neural runs---\eg, \citep{van2018ilps}---being more likely to uniquely retrieve a relevant document~\citep{Voorhees18} in comparison to traditional runs. If future neural runs, during reuse of the test collection, also had this property of finding previously unseen relevant results, then the evaluation of those new runs would be unfair, since no new judging is done during reuse. Although this indicates a potential problem, no previous work systematically analysed the reusability of test collections generated using traditional models towards evaluating the quality of such neural models.


%
%
%
%
%
%

%% file: 05-experiment.tex
\section{Experimental Analysis}
\label{sec:experiment}

We analysed the quality of test collections constructed using depth pooling from traditional vs. neural systems in terms of the number of relevant document identified, as well as in terms of the reusability of these pools based on the evaluation results obtained for systems of different types (neural vs. traditional systems). For this purpose, we use the data from The TREC Deep Learning Track~\citep{craswell2020overview}, details of which are described below.

%% file: 06-data.tex
\subsection{Task and datasets}
\label{sec:data}

The TREC Deep Learning Track has two tasks: Document retrieval and passage retrieval. Both tasks have large training sets based on human relevance assessments, derived from MS MARCO \citep{bajaj2016ms}. The test collection used in the track, which was generated using the depth-10 pools of the participating systems, contains $43$ queries. Judgments were done on a four-point scale: Perfectly relevant, highly relevant, relevant and irrelevant.
The track reported both NDCG@10 and MRR metrics, with NDCG@10 being the primary metric used in ranking the systems.

In total $10$ groups with a total of $38$ runs participated in the document retrieval task and $11$ groups with a total of $37$ runs participated in the passage retrieval task. For the document retrieval task, out of the $38$ runs, $27$ of them were based on neural models (models based on deep learning methods or use such models (e.g. BERT) as features)) and $11$ of them were based on traditional methods (models that are based on traditional, non-neural methods such as BM25). For the passage retrieval task, $26$ runs were based on neural models and $11$ of them were based on traditional methods. 

Top $10$ performing systems in both document ranking and passage ranking tasks were based on neural models. More details about the results of the Deep Learning Track can be found at the overview paper of the track~\citep{craswell2020overview}.

%% file: 07-result.tex
\subsection{Experimental Results}
\label{sec:result}

\subsubsection{Number of Relevant Documents Found}
Since the reusability of a test collection highly depends on the number of relevant documents identified, we first analysed the number of relevant documents identified if pools were to be constructed solely using (i) traditional runs vs. (ii) neural runs.

For this purpose, we divided the runs submitted to the Deep Learning Track into two categories: Traditional systems and neural systems. The runs were assigned to the two categories based on the original categorization used by the track, as described in Section~\ref{sec:data}. We then analysed the number of documents identified when test collections are constructed using depth-$k$ pooling by pooling top $k$ results from traditional systems vs. neural systems, for various cutoff values $k$.

\begin{figure}
\subfloat[Document retrieval task]{
  \includegraphics[width=0.5\linewidth]{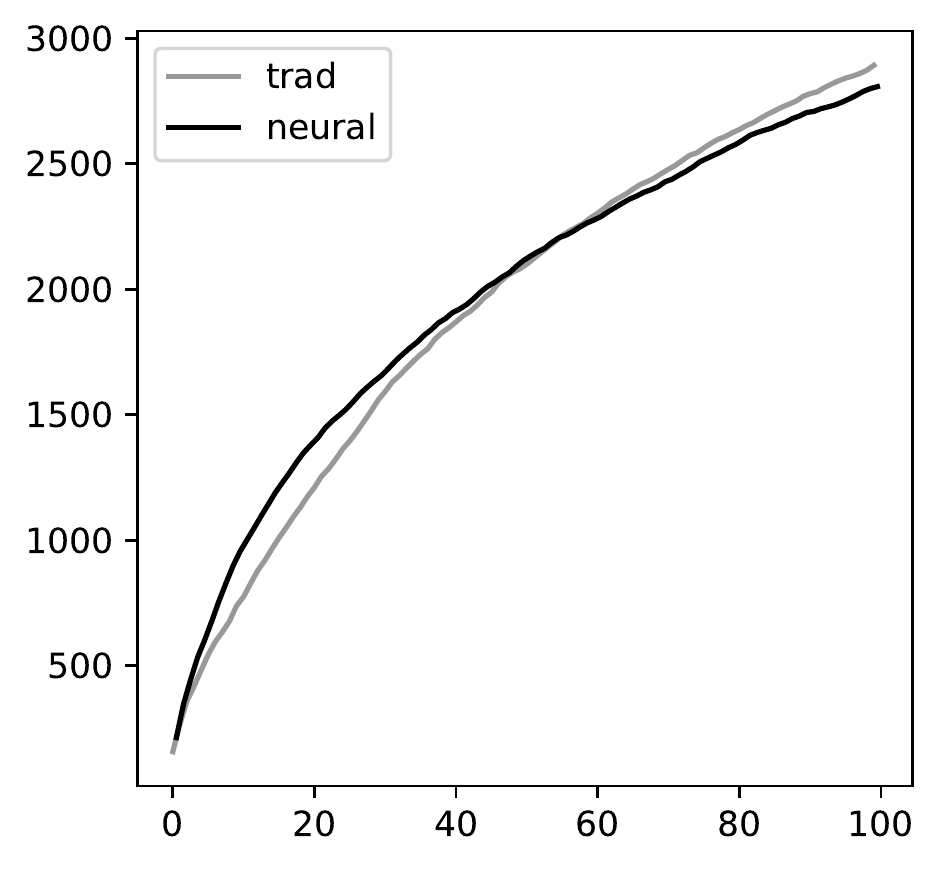}
}
\subfloat[Passage retrieval task]{
  \includegraphics[width=0.5\linewidth]{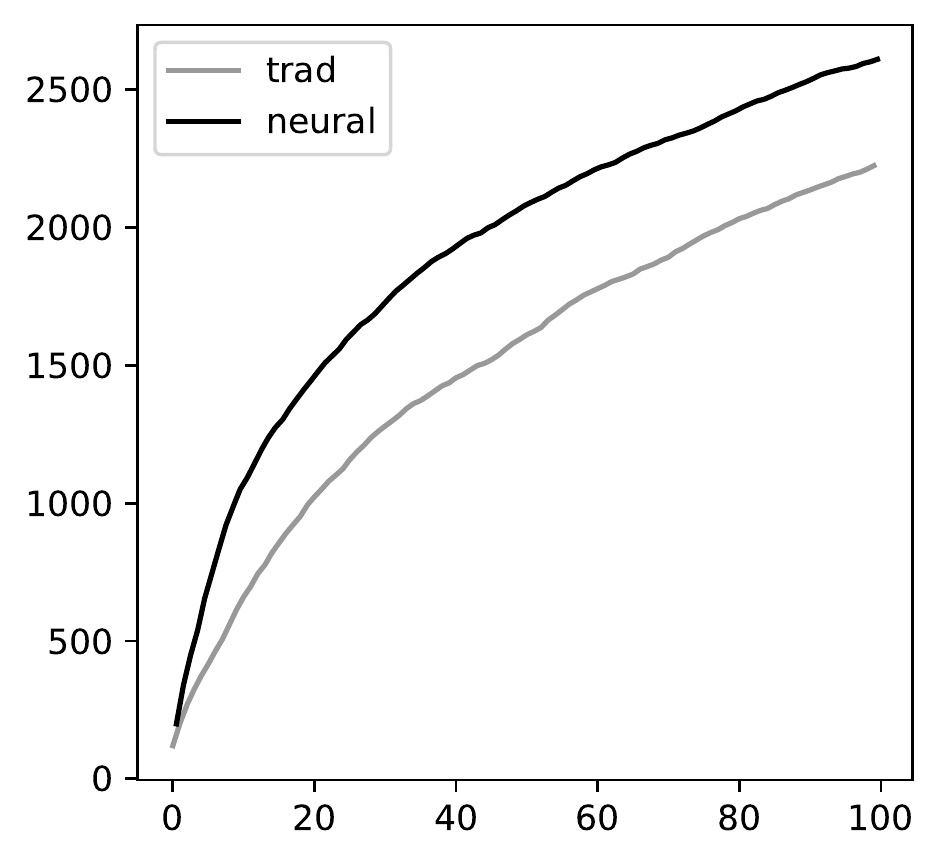}
}
\caption{Cumulative count of relevant results at each rank cutoff $k$. In document retrieval, neural runs find more relevant results at early cutoffs, but then are overtaken by traditional runs at later cutoffs. In passage retrieval, neural runs find more relevant results at all cutoffs.}\label{Fig_NumRels}
\vspace{-4ex}
\end{figure}

Figure~\ref{Fig_NumRels} shows the result of this experiment for the document retrieval task (left) and the passage retrieval task (right plot). The $x$ axis in the figures shows the cutoff value $k$ used to constructed the depth-k pools and the $y$ axis shows the number of relevant documents identified using pools constructed via traditional (grey line) vs. neural (black line) models. 

It can be seen that for both tasks neural models tend to find more relevant documents at early cutoff levels. For document retrieval task, neural runs seem to be overtaken by the traditional runs as one goes deeper in the ranking whereas for the passage retrieval task neural runs consistently find more relevant results at all cutoffs. Given that most IR metrics tend to be top heavy, these results raise concerns about the reliability of evaluation results in evaluating neural models with pools generated from traditional methods, a commonly faced scenario due to most existing test collections being generated solely using traditional models.

\subsubsection{Test Collection Reusability}

\begin{table}[t!]
\centering
\begin{tabular}{|l|c|c|c||c|c|c|}\hline
& \multicolumn{3}{c||}{MRR} & \multicolumn{3}{c|} {NDCG@10} \\ \hline \hline
Test System: & Trad & Neural & All & Trad & Neural & All \\ \hline
Trad Pool & 0.436 & -0.12 & -0.19 & 0.772 & 0.68 & 0.676 \\ \hline
Neural Pool & 0.769 & 0.635 & 0.842 & 0.774 & 0.836 & 0.852 \\ \hline
\end{tabular}
\caption{Average Kendall's tau correlations between actual metric and metric values computed using 10 randomly generated traditional (top row) vs. neural pools (bottom row) for document retrieval runs.}
\label{Tau_DocRanking}
\vspace{-6ex}
\end{table}

We then analysed the reusability of test collections generated via pooling top-$k$ results of systems of a particular type for evaluating systems that are of a different type, particularly focusing on traditional vs. neural system types. In particular, we are interested in the question as to whether pools generated using traditional systems can be reliably used to evaluate the quality of neural models, and vice versa. 

In order to evaluate the reusability of test collections generated using traditional pools towards evaluating the quality of neural models, we randomly split the traditional runs submitted to the TREC Deep Learning Track into two sets. We used the first set of systems to construct the test collection using depth-10 pooling (which we refer to as the \emph{traditional pool}), and we used the second set of systems together with the neural models as test systems, using which we analyse the reusability of the pools generated. Depth-10 pools were used in the pooling process since the original test collection for the Deep Learning Track was generated using depth-10 pooling.

In TREC, most groups tend to submit multiple runs and most of these runs tend to be different variants of the same system, which was also the case for the Deep Learning Track (as described in  Section~\ref{sec:data}). In order to avoid having a system in the test set that is very similar to a system used in constructing the pools, if one run from one group is randomly selected to be included in the pool, all the remaining runs from that group are also included in the pool.

We then used this test collection to evaluate the quality of test systems (neural systems, as well as the traditional systems that did not contribute to the pool). This way we can evaluate the reusability of the test collection constructed with traditional systems in terms of their fairness towards evaluating (i) the performance of neural systems within themselves, (ii) the performance of other traditional systems that did not contribute to the pool within themselves, and (iii) the relative performance of neural vs. traditional systems. 

Since the test collection for the Deep Learning Track was generated using depth-10 pooling, we also use such depth-10 pools evaluation results obtained by pooling all the systems submitted as our gold standard (which we refer to as the actual metric values). We then compare the actual metric values with metric values computed using the traditional pool (which we refer to as the estimated metric values). We then compute the Kendall’s tau correlation between the actual metric values and estimated metric values when (i) only traditional methods are used as the test systems, (ii) only neural models are used as the test systems, and (iii) systems of both types are used as the test systems. Since NDCG@10 and MRR were two of the primary metrics used in context of the Deep Learning Track, we also focused on these metrics as the primary evaluation metrics. 

Since the quality of the pools constructed could be highly affected by the type of runs that are randomly selected for constructing the pools, we repeat this process 10 times to construct 10 random pools generated by using different random splits of traditional runs, and compute the average Kendall’s tau correlation results over all the 10 runs. 

In order to evaluate the reusability of pools generated using neural models, we repeated the same procedure to by randomly selecting half of the neural runs for constructing the pools (which we refer to as neural pools), which are then used to evaluate the quality of neural models that did not contribute to the pool together with traditional models in a similar way as above.

\begin{table}[t!]
\begin{tabular}{|l|c|c|c||c|c|c|}\hline
& \multicolumn{3}{c||}{MRR} & \multicolumn{3}{c|} {NDCG@10} \\ \hline \hline
Test System: & Trad & Neural & All & Trad & Neural & All \\ \hline
Trad Pool & 0.63 & 0.004 & 0.0 & 0.789 & 0.574 & 0.612 \\ \hline
Neural Pool & 0.7 & 0.81 & 0.875 & 0.89 & 0.874 & 0.881 \\ \hline
\end{tabular}
\caption{Average Kendall's tau correlations between actual metric and metric values computed using 10 randomly generated traditional (top row) vs. neural pools (bottom row) for passage retrieval runs.}
\label{Tau_PassageRanking}
\vspace{-7ex}
\end{table}

Table~\ref{Tau_DocRanking} and Table~\ref{Tau_PassageRanking} show the average Kendall's tau values obtained over the 10 randomly constructed pools when pools are constructed using half of the traditional systems (upper row) and half of the neural systems (bottom row) for the document retrieval and passage retrieval tasks, respectively. The columns in the tables show the different types of test systems used in evaluation. It can be seen that for both document retrieval and passage retrieval tasks, traditional pools result in poor evaluation results for the neural systems. In fact, traditional pools seem to be worse than neural pools even for evaluating the quality of traditional systems that did not contribute to the pool! 

\begin{figure}[t!]
\centering
\subfloat[MRR Traditional Depth-10 Pool][MRR\\ Traditional Depth-10 Pool]{ \includegraphics[width=45mm]{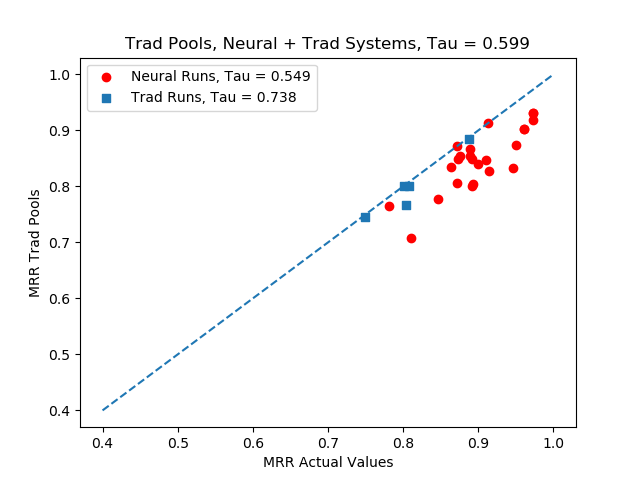}
}
\subfloat[MRR Neural Depth-10 Pool][MRR\\ Neural Depth-10 Pool]{
  \includegraphics[width=45mm]{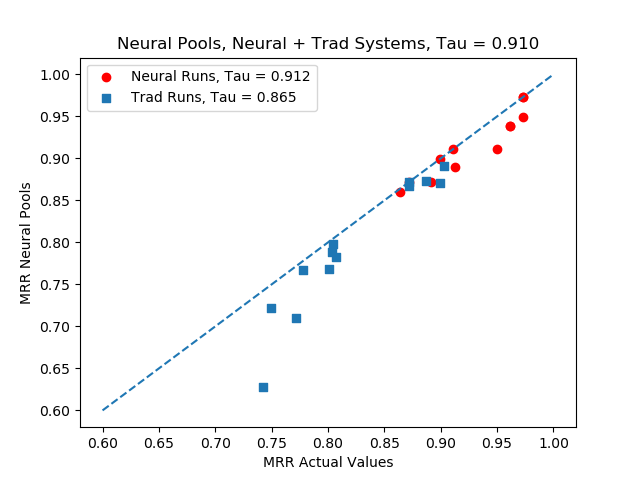}
}
\hspace{0mm}
\subfloat[NDCG@10 Traditional Depth-10 Pool][NDCG@10\\ Traditional Depth-10 Pool]{
  \includegraphics[width=45mm]{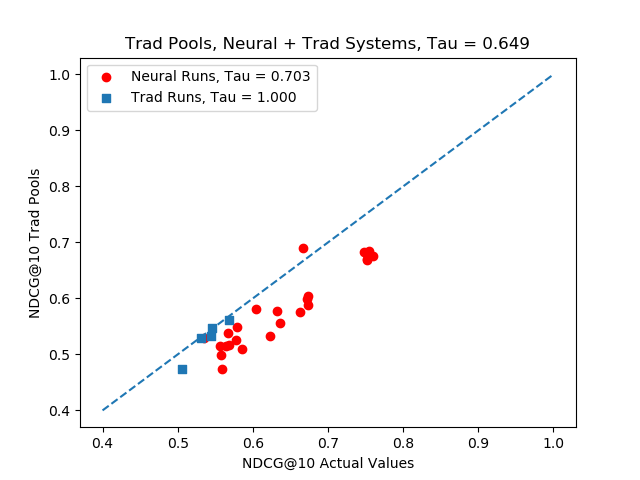}
}
\subfloat[NDCG@10 Neural Depth-10 Pool][NDCG@10\\ Neural Depth-10 Pool]{
  \includegraphics[width=45mm]{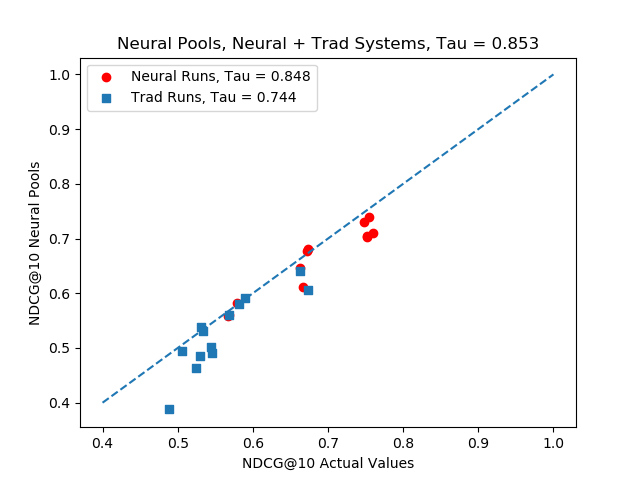}
}
\caption{MRR (top) and NDCG@10 (bottom) values for document retrieval task, when pools are generated using (left) traditional systems vs. (right) neural systems.}
\vspace{-4ex}
\label{Fig_DocRanks}
\end{figure}

Furthermore, in most cases the Kendall's tau correlation when all the test systems are considered seems to be less than the Kendall's tau scores for traditional systems and neural systems alone. This suggests that such pools are performing very poorly when the pairwise comparisons between the traditional vs. neural systems are considered. Hence, when a neural model is compared with a traditional model using a test collection generated via traditional pools, one might incorrectly infer that the neural model is performing worse than the traditional model. 

Note these findings could be partially related to the number and type of runs included in the pools. In our experiments, we were limited by the number and type of runs submitted to the Deep Learning Track. If more runs with more variety were used to create the pools, the resulting pools have the potential to result in more reliable evaluations of neural systems. However, the fact that the same exact pools Table~\ref{Tau_DocRanking} Table~\ref{Tau_PassageRanking} could lead to very different evaluation results in terms of their reliability when evaluating the quality of traditional vs. neural techniques is highly concerning. 

Out results suggest that existing test collections generated using traditional systems should be used with caution when evaluating the quality of neural models as the evaluation results obtained are likely to be unreliable and one might incorrectly infer that the quality of the neural run is worse than a baseline traditional or neural run. 

\begin{figure}[t!]
\subfloat[MRR Traditional Depth-10 Pool][MRR\\ Traditional Depth-10 Pool]{
  \includegraphics[width=45mm]{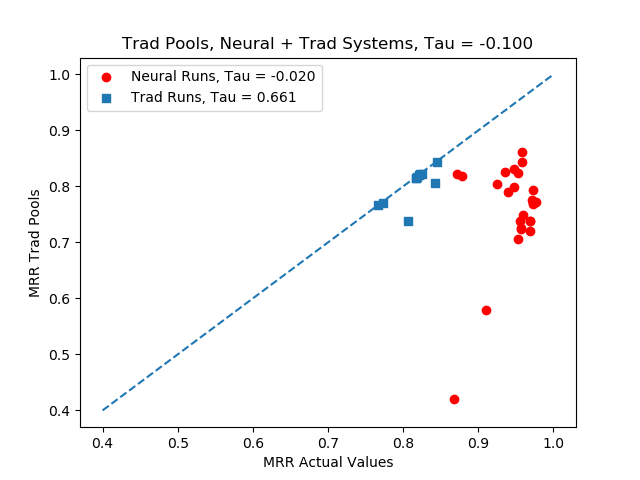}
}
\subfloat[MRR Neural Depth-10 Pool][MRR\\ Neural Depth-10 Pool]{
  \includegraphics[width=45mm]{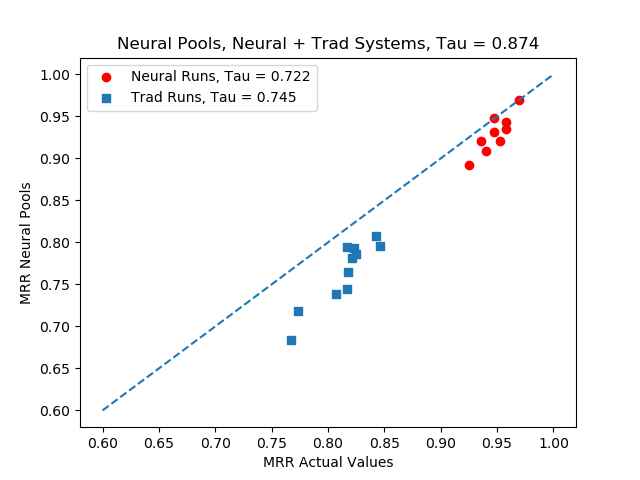}
}
\hspace{0mm}
\subfloat[NDCG@10 Traditional Depth-10 Pool][NDCG@10\\ Traditional Depth-10 Pool]{
  \includegraphics[width=45mm]{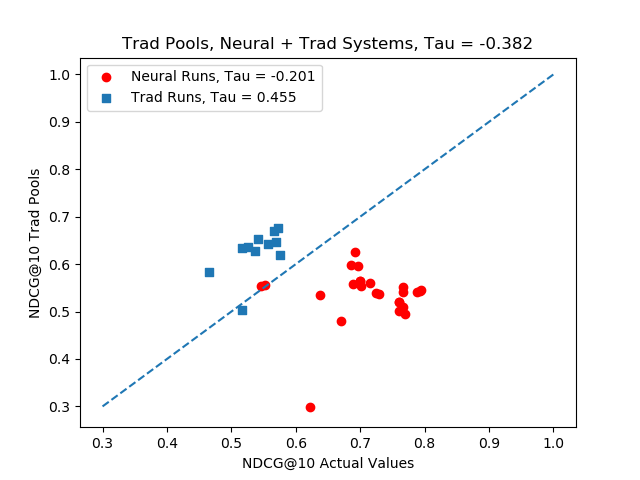}
}
\subfloat[NDCG@10 Neural Depth-10 Pool][NDCG@10\\ Neural Depth-10 Pool]{
  \includegraphics[width=45mm]{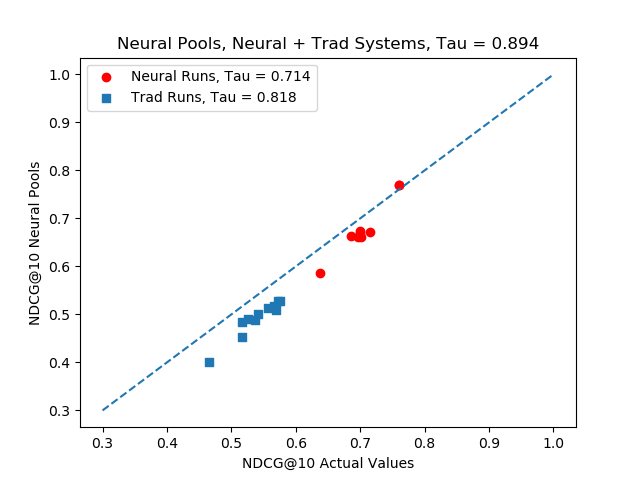}
}
\caption{MRR (top) and NDCG@10 (bottom) values for passage retrieval task, when pools are generated using (left) traditional systems vs. (right) neural systems.}
\vspace{-3ex}
\label{Fig_PassageRank}
\end{figure}

Figure~\ref{Fig_DocRanks} and Figure~\ref{Fig_PassageRank} show how such evaluations look like in detail for a randomly picked pool for the document retrieval and passage retrieval tasks, respectively. The $x$ axis in the plots show the actual metric values when all the systems are used to generate the pools and the $y$ axis shows the estimated metric values computed when half of the traditional (left plots) or neural systems (right plots) are used to generate the pools. The plots also contain line $y=x$ for comparison purposes. The titles in the plots show the Kendall’s tau correlation between the actual and the estimated metric values when all systems are considered in the test set. The plots also show the Kendall’s tau correlation values within the neural models, as well as within the traditional models in the test set. 

It can be seen that pools generated using the traditional pools are particularly unreliable in evaluating neural runs and may have a tendency to underestimate the quality of the neural runs, whereas pools generated via the neural runs tend to be more reliable for evaluating the quality of both traditional and neural systems.

%% file: 09-conclusion.tex
\section{Conclusion}
\label{sec:conclusion}

We analyse the reusability of test collections when they are created solely using systems of a particular type (e.g. traditional systems based on BM25, language modelling, etc.) towards evaluating the quality of systems that are of a different type (e.g. neural systems based on deep learning models or use such models as features). 

Our results demonstrate that evaluation results obtained using test collections that are created solely using traditional runs are not very reliable in terms of evaluating the quality of neural systems. In particular, our findings suggest that such test collections should be used with caution when evaluating the quality of neural systems as they may lead to incorrect conclusions regarding how the quality of a neural model compares with a traditional model, as well as how the neural model compares with another baseline neural model. 

While our results suggest that pools systematically generated using one system type (e.g. traditional systems) may be problematic when they are used to evaluate the quality of systems of a different type (e.g. neural systems), reusability of test collections generated using neural models towards other (unseen) neural models is a question we would like to analyse further in the future. 